\documentclass[]{llncs}
\usepackage{graphicx}
\pdfoutput=1
\usepackage[utf8x]{inputenc}
\usepackage{epstopdf}
\usepackage{algorithm,algorithmic}
\usepackage{complexity}
\usepackage{amsmath,amssymb,amsfonts}
\usepackage{textcomp}
\usepackage{xcolor}
\begin{document}
\title{ A Note on Hardness of Multiprocessor Scheduling with Scheduling Solution Space Tree }
%
%
\author{Debasis Dwibedy  \and
Rakesh Mohanty}
\authorrunning{D. Dwibedy and  R. Mohanty}
%
\institute{Veer Surendra Sai University of Technology, Burla, 768018, Odisha, India
\email{\{debasis.dwibedy, rakesh.iitmphd\}@gmail.com}}
%
\maketitle              
\begin{abstract}
We study the computational complexity of the non-preemptive scheduling problem of a list of independent jobs on a set of identical parallel processors with a makespan minimization objective. We make a maiden attempt to explore the combinatorial structure showing the exhaustive solution space of the problem by defining the \textit{Scheduling Solution Space Tree (SSST)} data structure. The properties of the \textit{SSST} are formally defined and characterized through our analytical results. We develop a  unique technique to show the problem $\mathcal{NP}$ using the SSST and the \textit{Weighted Scheduling Solution Space Tree (WSSST)} data structures. We design the first non-deterministic polynomial-time algorithm named \textit{Magic Scheduling (MS)} for the problem based on the reduction framework. We also define a new variant of multiprocessor scheduling by including the user as an additional input parameter. We formally establish the complexity class of the variant by the reduction principle. Finally, we conclude the article by exploring several interesting open problems for future research investigation.  
\keywords{Combinatorial Structure \and Computational Complexity \and Hardness \and Makespan \and Multiprocessor Scheduling \and Multiuser \and $\mathcal{NP}$-completeness \and Nondeterministic Algorithm \and Reduction \and Scheduling Solution Space Tree}
\end{abstract}
\section{Introduction} \label{subsec: Introduction}
The non-preemptive \textit{Multiprocessor Scheduling Problem (MPSP)} deals with the scheduling of a list of jobs on a set of identical parallel processors to minimize the completion time of the job that finishes last in the schedule, i.e., makespan \cite{McNaughton59}. The MPSP is a typical combinatorial minimization problem where with the increase in the number of jobs and processors, the scheduling decision leads to exponential solution space. For example, in the case of a list of $n$ jobs and $m$ processors, we have at most $m^n$ feasible schedules for the job list. In many scientific studies, authors have attempted to explore optimal scheduling that incurs the minimum value of makespan. These studies generally have ended up proving the problem $\mathcal{NP}$-complete providing a good indication of the hardness of the problem. Nevertheless, the current literature lacks an optimal scheduling algorithm that runs in polynomial time in the length of jobs or processors. The following natural questions that arise here are :
\begin{itemize}
\item what makes the MPSP problem harder to solve in polynomial time even for $m=2$?
\item how do we formally represent the combinatorial structure of the problem? 
\item how does the solution space affect the complexity class? 
\end{itemize}
In our current study, we address the above questions. The idea is clear, if at all the problem is yet to be solved efficiently, we must develop a strategy to explore the intrinsic details such as the exhaustive solution space of the problem so that we can precisely define the complexity class and in the future, we would possibly be able to design an optimal scheduling algorithm. Another perspective of our study is related to the $\mathcal{NP}$ proof of the MPSP problem. The first problem proved $\mathcal{NP}$-complete was the $3$-SAT problem by Cook \cite{Cook70}. The author exclusively proved that $3$-SAT$\in \mathcal{NP}$ and $3$-SAT$\in \mathcal{NP}$-hard. Subsequently, several complex combinatorial problems have been shown $\mathcal{NP}$-complete by the method of reduction from the well-known $\mathcal{NP}$-complete problems with less concern towards the $\mathcal{NP}$ proof of these problems. Although many researchers have shown the variants of MPSP $\mathcal{NP}$-complete, there is hardly any attempt to exclusively develop a proof technique for MPSP$\in \mathcal{NP}$. The design of a model or a non-deterministic polynomial-time algorithm for MPSP would possibly pave the way for the development of a deterministic polynomial-time algorithm for all $\mathcal{NP}$-complete problems.\\\\
\textbf{Our Contribution.} We develop the \textit{Scheduling Solution Space Tree (SSST)} to explore the combinatorial structure with exhaustive solution space of \textit{MPSP}. The properties of the \textit{SSST} are formally defined and characterized through our analytical results. We develop a  unique technique to show that MPSP $\in \mathcal{NP}$ by mapping the construction of the SSST to a non-deterministic Turing Machine that can verify a given \textit{scheduling solution} as a certificate in polynomial time. The SSST is shown as a polynomial-time verifier with its variant named Weighted Scheduling Solution Space Tree (WSSST) by following an interactive proof method. We prove that MPSP is $\mathcal{NP}$-hard by an alternate yet simple reduction technique. We make a maiden attempt to design a non-deterministic polynomial-time algorithm named \textit{Magic Scheduling (MS)} for the problem based on the reduction framework. A variant of MPSP is defined by considering the \textit{user} as an additional input parameter and named the problem as \textit{Multiuser Multiprocessor Scheduling Problem (MUMPSP)}. We prove the complexity class of MUMPSP by showing an equivalence between MPSP and MUMPSP.\\\\
\textbf{Organization.} We organize the rest of the paper as follows. Section 2 discusses the foundation of complexity classes and highlights the state-of-the-art literature on the related hardness studies of the variants of the MPSP problem. Section 3 formally defines and characterizes the \textit{SSST} data structure based on the MPSP problem. Here, we explore the combinatorial structure and derive some interesting results. Section 4 highlights the computational hardness of the MPSP. Here, we propose a general reduction framework and design a non-deterministic polynomial-time algorithm named \textit{Magic Scheduling (MS)} for the MPSP problem. We show the computational hardness of MPSP by our proposed SSST and algorithm MS with new analytical results. Section 5 formally defines the MUMPSP problem and proves its complexity class. Section 6 highlights some non-trivial open problems for future work.
\section{Background and State-of-the-art Literature}
Here, we highlight and discuss the fundamental aspects of the complexity classes for a basic understanding. We also overview the state-of-the-art literature on the related hardness studies of the variants of the MPSP problem.
\subsection{Complexity Classes: Preliminaries}
\begin{figure}[]
\centering
\includegraphics[scale=0.65]{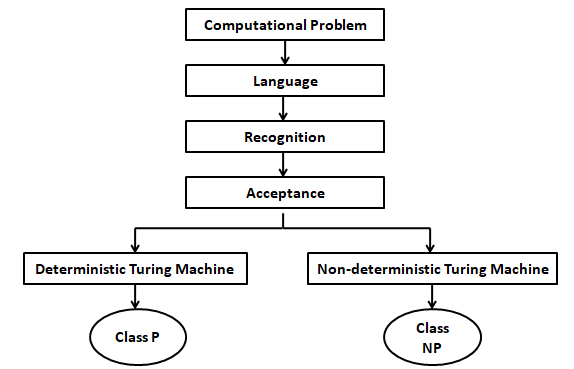}
\caption{Foundation of Complexity Classes}
\label{foundationofcomplexityclasses}
\end{figure}

The complexity defines the hardness of solving a computational problem. A complexity class identifies a set of problems that are similar in hardness. Cook \cite{Cook70} and Karp \cite{Karp72} were the first to define the basic complexity classes such as $\mathcal{P}$, $\mathcal{NP}$. They also established the formal relationship between $\mathcal{P}$ and $\mathcal{NP}$ in terms of language recognition by the Turing Machine ($\mathcal{TM}$).
We transform the fundamental idea of defining the complexity classes in Fig \ref{foundationofcomplexityclasses} for simplicity in understanding.
 \\
We can express any computational problem $X$ as a formal language $L$, where $L\subseteq \sum^{*}$ and the $\sum^{*}$ is a set containing all strings over any finite alphabet \cite{Cook70}. Furthermore, we can transform a language into a string recognition problem for the $\mathcal{TM}$ \cite{Karp72}. 
Since the $\mathcal{TM}$ can either accept/reject a given string $w\in L$, we create an instance of $X$ and map to an equivalent string $w^{'}$ to yield a yes/no answer \cite{Karp72}. The acceptance of $w^{'}$ implies a 'yes' answer, and if it is realized by a deterministic $\mathcal{TM}$ in polynomially bounded time with the length of $w^{'}$, then $P_1$ is considered to be in $\mathcal{P}$ class \cite{Cook70}, \cite{Karp72}, \cite{Aho74}. However, if $w^{'}$ is accepted only by a non-deterministic $\mathcal{TM}$ in polynomial time, then $X$ belongs to $\mathcal{NP}$ class \cite{Cook70}. By the polynomial time, we mean the time taken for the computation is in the form of $n^c$, where $n$ is the length of the input and $c (\geq 1)$ is any positive number \cite{Aho74}. Informally a non-deterministic algorithm for $X$ can be considered equivalent to that of the corresponding language recognition algorithm for the non-deterministic $\mathcal{TM}$ \cite{Karp72}.\\
Since a $\mathcal{TM}$ can either accept/reject a given string, we have to represent a computational problem as a language recognition problem that requires a yes/no answer. The answers 'yes' and 'no' correspond to acceptance and rejection of a language. We usually refer to the \textit{decision version} of a problem as a language recognition problem that requires a yes/no answer \cite{Aaronson16}. Therefore, to show a problem $\mathcal{NP}$, we generally state its decision version and verify the existence of a solution with a value equal to some \textit{threshold} (mostly the optimal value). \\
Concerning the combinatorial structure of MPSP, we do not require rigorous formal definitions of the complexity classes. In general, we can define the class $\mathcal{P}$, consisting of a set of problems for which polynomial-time algorithms exist, whereas the class $\mathcal{NP}$ covers a set of problems for which non-deterministic polynomial-time algorithms exist. Alternatively, we can define a problem that belongs to the $\mathcal{NP}$ class if the solution to a given instance of a problem is verifiable in polynomial time.
We can say clearly that $\mathcal{P} \subset \mathcal{NP}$ but the question remains open whether $\mathcal{P} = \mathcal{NP}$ \cite{Aaronson16}.\\
In a milestone paper \cite{Cook70}, Cook introduced the $\mathcal{NP}$-complete theory by taking into account the \textbf{Satisfiability (SAT)} problem. We can formally define the SAT problem as follows.\\\\
\textit{Given a collection of $n$ clauses $C_1, C_2, \ldots, C_n$, where each clause $C_i$ is a disjunction of the literals from a set $U=\{u_1, u_2, \ldots, u_t, \bar{u_1}, \bar{u_2}, \ldots, \bar{u_t}\}$. Is $\bigwedge_{i=1}^{n}{C_i}$ satisfiable, i.e., is there a subset $U^{'}\in U$ such that there exists no $C_i$ with complementary pair of literals $(u_j, \bar{u_j})$ in $U^{'}$, and $U^{'}\cap C_i\neq \phi$, $\forall_i$?}\\\\
The SAT problem, where each clauses uses three literals is called the \textbf{$3$-SAT} problem.\\\\
\textbf{Theorem 2.1} \cite{Cook70}. \textit{$3$-SAT is $\mathcal{NP}$-complete.}\\\\
Cook \cite{Cook70} proved the $\mathcal{NP}$-completeness of the $3$-SAT
problem by introducing the reduction technique that maps any instance of a problem $X\in \mathcal{NP}$ to an equivalent boolean expression in CNF, i.e., conjunctive normal form. Thus a polynomial-time algorithm for the $3$-SAT problem could be employed to design a polynomial-time algorithm for any problem $X^{'}\in \mathcal{NP}$. This implies, $\mathcal{P}=\mathcal{NP}$ if and only if $3$-SAT$\in \mathcal{P}$.\\
Formally, a problem $X_1$ is considered to be $\mathcal{NP}$-complete, if $X_1 \in \mathcal{NP}$ and there exists a function $f$, which can \textit{reduce}($\propto$) in polynomial time an instance $x$ of any problem $X_2 \in \mathcal{NP}$ to at least an instance $y$ of $X_1$ such that $y=f(x)$ \cite{Garey90}. Subsequently we can claim that there exists a polynomial time solution for $y$ if and only if $x$ has a solution in polynomial time. \\\\
If $X_1$, $X_2$, and $X_3$ are three $\mathcal{NP}$-complete problems, then they hold the following properties.
\begin{itemize}
\item Property 1. If $X_2\propto X_1$ and $X_1\propto X_3$, then $X_2\propto X_3$.
\item Property 2. Let the problem $A^{'}$ is the special case of problem $A$ and if $A^{'}\in \mathcal{NP}$, \hspace*{1.9cm}then  $A\in \mathcal{NP}$ and if $A^{'}\in \mathcal{NP}$-complete then $A\in \mathcal{NP}$-complete.
\end{itemize}
Karp \cite{Karp72} established the reducibility theory by showing several combinatorial optimization problems $\mathcal{NP}$-complete using the novel reduction technique. We highlight Karp's approach of proving a new problem $\mathcal{NP}$-complete by Lemma 2.2.\\\\
\textbf{Lemma 2.2} \cite{Karp72}. Let $X_1$ and $X_2$ be two computational problems. If $X_1, X_2 \in \mathcal{NP}$, $X_1\in \mathcal{NP}$-complete, and $X_1\propto X_2$, then $X_2\in \mathcal{NP}$-complete.\\\\
\textbf{Proof} We are required to show that $X_2\in \mathcal{NP}$-complete. Since $X_2 \in \mathcal{NP}$, we only need to show that for every problem $X_3\in \mathcal{NP}$, $X_3\propto X_2$. We are given that $X_1\in \mathcal{NP}$-complete, implies  $X_3\propto X_2$. Already we know that $X_1\propto X_2$. Therefore by Property 1, we have $X_3\propto X_2$. \hfill\(\Box\)\\\\
The $\mathcal{NP}$-completeness theory of Karp \cite{Karp72} rejuvenated the non-trivial question of whether $\mathcal{P}=? \mathcal{NP}$. Despite some quality efforts from many intelligent researchers for the last five decades, no polynomial-time algorithms have been explored for the $\mathcal{NP}$ class problems to date. It seems that $P = \mathcal{NP}$ is quite unlikely. Therefore, we presumably consider that $\mathcal{P}\subset \mathcal{NP}$.  
For an exhaustive read of the rigorous mathematical definitions, properties, and in-depth analysis of the complexity classes, we refer the readers to the following outstanding articles \cite{Cook70}, \cite{Karp72}, \cite{Aho74}, \cite{Garey90}, \cite{Aaronson16}.\\\\
The reducibility theory has put the foundation for locating the borderline between the $\mathcal{P}$ (easy) and $\mathcal{NP}$-complete (hard) problems. The precise understanding of the borderline helps to figure out the parameters of a problem that determine its complexity class and also assists in formulating effective solution methods. For example, proving a problem $\mathcal{NP}$-complete establishes a formal argument to use enumerative solution methods such as approximation, local search, branch, and bound. Before presenting our results on the hardness of the MPSP problem, we first overview the related studies in the next section. 
\subsection{Overview of the State-of-the-art Related Work}
In this section, we review the state-of-the-art literature on the hardness study of the scheduling problem for identical machine settings with the optimality criteria, such as makespan, the sum of completion times, weighted sum of completion times, and weighted mean completion time.\\  Karp \cite{Karp72} first proved that the problem of scheduling a list of jobs on a set of parallel processors to complete all jobs within a given deadline is $\mathcal{NP}$-complete. The author showed the proof by a polynomial-time reduction of an instance of the problem from an instance of the well-known \textit{0/1- Knapsack problem}. Bruno et al. \cite{Bruno74} proved that scheduling a list of independent jobs to minimize the \textit{weighted mean completion time (wmct)} of the job schedule is $\mathcal{NP}$-complete. The result was shown by a polynomial-time reduction from the $0/1$-Knapsack problem. Garey and Johnson \cite{Garey74} studied the hardness of the scheduling problem with $m(\geq 2)$-identical processors, where each job $J_i$ follows a partial order and requires some resources for its execution. The authors proved that the problem is $\mathcal{NP}$-complete even for $m=2$ and $n$ jobs by a polynomial-time reduction from the well-studied \textit{Node Cover} problem. The variant of the problem with $m=5$ and $n=8$ was shown $\mathcal{NP}$-complete by a polynomial-time reduction from the well-known \textit{$3$-Dimensional Matching problem}. Ullman \cite{Ullman75} studied a variant of the scheduling problem proposed by Garey and Johnson in \cite{Garey74} with $m=2$ to minimize the makespan, where the processing time $p_i=\{1, 2\}$, $\forall J_i$. The author showed the $\mathcal{NP}$-completeness of the problem by a polynomial-time reduction from the \textit{$3$-SAT} problem. The readers can find the comprehensive survey of the seminal contributions, and early results on the computational hardness of the multi-processor scheduling problem and its variants in \cite{Lenstra77}.\\\\
Recently, there have been increasing interests in the study of the parameterized complexity classes of variants of the MPSP problem. The idea of such a class is to consider exponential-time algorithms for solving $\mathcal{NP}$-complete problems while restricting them to a smaller parameter. Generally, an instance of any parameterized problem $X$ consists of two input parameters, i.e., $x$ and $k$. The problem $X$ is Fixed Parameter Tractable (FPT), if $X$ is solvable in time $f(k)\cdot poly(|x|)$ for any computable function $f$. The problem $X$ is $W[t]$-hard, if there exists a function $f$, which can reduce $(\propto)$ in time $f(k)\cdot poly(|x|)$ an instance $(x^{'}, k^{'})$ of any problem $X^{'}\in W[t]$-hard to at least an instance $(x, k)$ of $X$ such that $k\leq g(k^{'})$ and $(x^{'}, k^{'})\in X^{'} \iff (x, k)\in X$, where $f$ and $g$ are two computable functions. There is a hierarchy of parameterized complexity classes, i.e., $FPT \subseteq W[1] \subseteq W[2] \subseteq \ldots \subseteq W[P]$. Bodlaender and Fellows \cite{Bodlaender95} showed the precedence-constrained deadline-based $m$-processor scheduling problem $W[2]$-hard by a polynomial-time reduction from the parameterized \textit{Dominating Set} problem. Mnich and Wiese \cite{Mnich15} investigated the problem of scheduling a list $A=\{J_1, J_2, J_3, \ldots, J_n\}$ of $n$ jobs on a single processor with  a flexibility of rejecting a set $A^{'}\subseteq A$ of at most $r$ jobs. Each job $J_i$ is associated with a weight $w_i$, processing time $p_i$, and each rejected one incurs a rejection cost of $e_i$. The objective is to minimize $\sum_{J_i\in A-A^{'}}^{}{w_ic_i} + \sum_{J_i\in A^{'}}^{}{e_i}$, where $c_i$ is the completion time of job $J_i$. The authors proved that the problem is $W[1]$-hard for the parameter $r$ of number of rejected jobs by a polynomial-time reduction from the well-known parameterized \textit{$r$-Subset Sum} problem.   Bevern et al. \cite{Bevern16} proved that the problem is $W[2]$-hard with respect to the width of the partial order of the jobs. The authors showed their claim by a polynomial-time reduction from the parameterized \textit{Shuffle-product} problem.\\
Blazewicz et al. \cite{Blazewicz01} proved that the problem of scheduling a list of unit-sized coupled jobs on a single processor with precedence constraints and makespan minimization objective is $\mathcal{NP}$-hard. A coupled job consists of two operations, where the execution of the second operation starts only after the completion of the first one. The authors showed the $\mathcal{NP}$-hardness proof by a polynomial-time reduction from the well-known \textit{Balanced-colouring of Graphs with Partially-ordered Vertices} problem. Garg et al. \cite{Garg07} investigated the problem of scheduling a list of jobs on a set of $m$-processor to minimize the sum of jobs completion times, where each job consists of components of different types with unit size, and each machine is capable of processing components of a single type. The authors proved the problem $\mathcal{NP}$-hard by a polynomial-time reduction from the well-studied \textit{Vertex Cover} problem. Ambuhl et al. \cite{Ambuhl11} studied the problem of scheduling a list of jobs on $1$-processor with precedence constraints to minimize the weighted sum of jobs completion times, where each job has a non-negative weight. The authors proved that the problem is $\mathcal{NP}$-hard by a polynomial-time reduction from the vertex cover problem.
Svensson \cite{Svensson11} studied the problem of scheduling a list of jobs on $m$-processor with precedence constraints to minimize the makespan. The author established a unique relationship between the considered problem and the problem of scheduling a list of weighted jobs on a single processor with precedence constraints to minimize the sum of weighted jobs completion times. The author proved that if the $1$-processor case is computationally hard to approximate closer to a factor of $2-\delta$, then the $m$-processor problem even for unit-sized jobs is also harder to approximate within a factor of $2-\epsilon$, where $\epsilon \rightarrow 0$ as $\delta \rightarrow 0$. \\
Bellenguez et al. \cite{Bellenguez15} studied the preemptive scheduling of a list of jobs on $3$-processor to minimize the sum of jobs completion times, where each job has a release time. The authors showed the $\mathcal{NP}$-hardness proof by a polynomial-time reduction from the $3$-Partition problem. Zhang et al. \cite{Zhang18} investigated the problem of scheduling a list of jobs on $1$-processor to minimize the makespan with time restrictions, and the following $K$-constraint. Given, $K=2$, for any real $t$, no unit-sized time interval $[t, t+1)$ is permitted to intersect greater than $K$ jobs. The authors proved the problem $\mathcal{NP}$-hard by a polynomial-time reduction from the well-known \textit{Equal Cardinality Partition} problem. Davies et al. \cite{Davies21} studied the problem of scheduling a list of jobs on $m$-processor with precedence constraints to minimize the makespan. For each pair of jobs satisfying the precedence relation, there exists a communication delay, i.e., the duration of time the scheduler waits between the two dependent jobs if they are assigned to different machines. The author showed the $\mathcal{NP}$-hardness proof of the problem by a reduction from the well-known \textit{Unique Machines Precedence-constrained Scheduling} problem.\\
It is evident from the literature that the hardness proofs of all $\mathcal{NP}$-complete scheduling problems offer a polynomial-time reduction from well-known $\mathcal{NP}$-complete problems by following \textit{Property 1} of the complexity classes. However, the existing literature lacks a formal method to show explicitly that \textit{MPSP} is $\mathcal{NP}$. The $\mathcal{NP}$-hardness of \textit{MPSP} was shown in \cite{Karp72}, \cite{Ullman75} by a reduction from the \textit{Partition problem} however, the result shown was intuitive and required refinement for the ease of understanding from a beginner's perspective. On this note, we present a systematic study to explore the hardness of the MPSP problem. We also prove the complexity class of the MUMPSP problem. We present a summary of well-known related hardness studies in Table \ref{Related Hardness Studies}.    
\begin{table}
\centering
\caption{Well-known Studies on $\mathcal{NP}$-completeness of the Multiprocessor Scheduling Problem }
\begin{tabular}{cp{7.3cm}p{3.3cm}}
\hline
\textbf{Year, Author(s)} & \textbf{Variants of MPSP} & \textbf{Reduction}\\
\hline
1972, Karp \cite{Karp72}  & $2$-processor scheduling with deadline & $0/1$-Knapsack\\
1974, Bruno et al. \cite{Bruno74} & $m$-processor scheduling with weighted mean completion time & $0/1$-Knapsack\\
1974, Garey, Johnson \cite{Garey74} & $2$-processor resource-constrained scheduling  & Node Cover\\
1975, Ullman \cite{Ullman75} & $2$-processor resource-constrained scheduling & $3$-SAT \\
1977, Lenstra et al. \cite{Lenstra77} & $2$-processor scheduling with makespan minimization & $2$-Partition\\
1995, Bodlaender, Fellows \cite{Bodlaender95} & Precedence-constrained $m$-processor scheduling problem is $W[2]$-hard for parameterized complexity class & Dominating Set\\
2001, Blazewicz J et al. \cite{Blazewicz01} & Scheduling of unit-sized coupled jobs on $1$-processor with precedence-constrained and makespan minimization & Balanced-colouring of graphs with partially-ordered vertices\\
2007, Garg et al. \cite{Garg07} & $m$-processor scheduling to minimize the sum of completion times & Vertex Cover\\
2011, Ambuhl et al. \cite{Ambuhl11} & $1$-processor scheduling with precedence constraints to minimize the weighted sum of the completion times & Vertex Cover \\
2011, Svensson \cite{Svensson11} & Precedence constrained $m$-processor scheduling with makespan minimization & Precedence-constrained $1$-processor scheduling with sum of weighted completion times \\
2015, Mnich, Wiese \cite{Mnich15} & Scheduling on $1$-processor with job rejection to minimize weighted sum of completion times is $W[1]$-hard & $r$-Subset Sum\\
2015, Bellenguez et al. \cite{Bellenguez15} & $3$-processor preemptive scheduling with release date to minimize the sum of completion time & $3$-Partition \\
2016, Bevern et al. \cite{Bevern16} & The precedence-constrained scheduling is $W[2]$-hard w.r.t. the width of the partial order & Shuffle-product\\
2018, Zhang et al. \cite{Zhang18} & $1$-processor scheduling with time restriction to minimize the makespan & Equal Cardinality Partition \\
2021, Davies et al. \cite{Davies21} & $m$-processor scheduling with non-uniform delay & Unique Machines Precedence-constrained Scheduling \\
\hline  
\end{tabular}
\label{Related Hardness Studies}
\end{table}
\section{Combinatorial Structure of Multiprocessor Scheduling Problem}
Here, we first formally define the MPSP problem, then present its combinatorial structure by proposing the SSST data structure, followed by a discussion on our analytical results.\\\\
\textbf{Multiprocessor Scheduling Problem (MPSP)}
\begin{itemize}
\item Inputs: \begin{itemize}
\item Given, $M=\{M_1, M_2, M_3, \ldots, M_m\}$ is the set of $m$ identical processors, and $J=\{J_1, J_2, J_3, \ldots, J_n\}$ is the list of $n$ jobs, where $n\ggg m$. 
\item Processing time $p_i$ for job $J_i$, where $p_i\geq 1$ and $1\leq i \leq n$.
\end{itemize}
\item Output: A schedule $S=\{J^1, J^2, \ldots, J^m\}$ is the collection of $m$ disjoint partitions of set $J$ such that $\bigcup_{j=1}^{m}{J^j}=J$, and for each pair of partitions $(J^a, J^b)\in S$, we have $J^a \cap J^b=\phi$, where $a\neq b$ and each $J^j$($\subseteq J$) represents a set of jobs assigned to respective machine $M_j$.
Each $M_j$ has a load $l_j$, where $l_j=\sum_{J_i\in M_j}^{}{p_i}$. The output parameter is \textit{makespan}, which is represented as $C_{max}$, where $C_{max}=\max_{1\leq j \leq m}^{}{l_j}$. 
\item Objective: To minimize $C_{max}$.
\item Assumptions: \begin{itemize}
\item Independent jobs: The set $J$ is free from the partial order ($\prec$) relation. Thus, jobs can execute in parallel.
\item Non-preemption: If a job $J_i$ with processing time $p_i$ starts its execution on any $M_j$ at time $t$, then it continues on $M_j$ till time $t+p_i$.
\end{itemize}
\end{itemize}
We define two new characterizations of the \textit{Schedule} $S$ such as \textit{Partial Schedule} ($S_p$) and \textit{Essential Schedule ($S_e$)} as follows.\\\\  
\textbf{Definition 3.1. Partial Schedule ($S_p$)} is the collection of $m$ disjoint partitions of a set $J'\subset J$ such that $\bigcup_{j=1}^{m}{J'^{j}}$=$J'$, where $J'^{j}\subseteq J'$,  and for each pair of partitions ($J^a, J^b$)$\in S_p$, we have $J^a \cap J^b=\phi$, where $1\leq a, b\leq m$ and $a\neq b$.\\\\
\textbf{Definition 3.2. Essential Schedule ($S_e$)} is a Schedule $S$, where $|J^j|\geq 1$, \hspace*{0.3cm} $\forall{j}$.

\subsection{Combinatorial Structure of MPSP as Scheduling Solution Space Tree}
\begin{figure}[]
\centering
\includegraphics[scale=0.5]{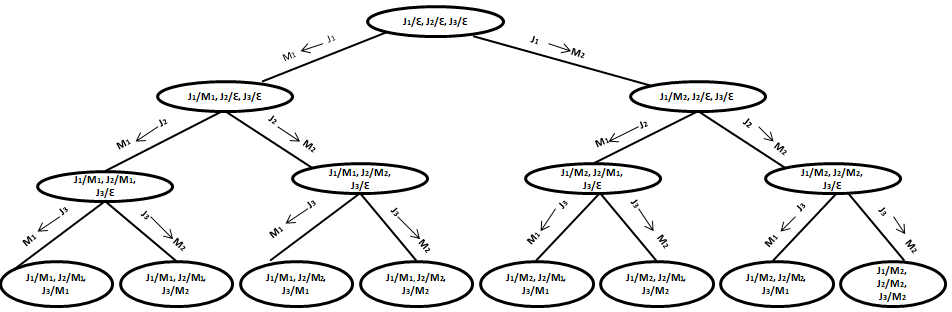}
\caption{Scheduling Solution Space Tree for 3 Jobs and 2 Processors}
\label{Schedulingsolutionspacetree}
\end{figure}

\begin{figure}[!htbp]
\includegraphics[scale=0.5]{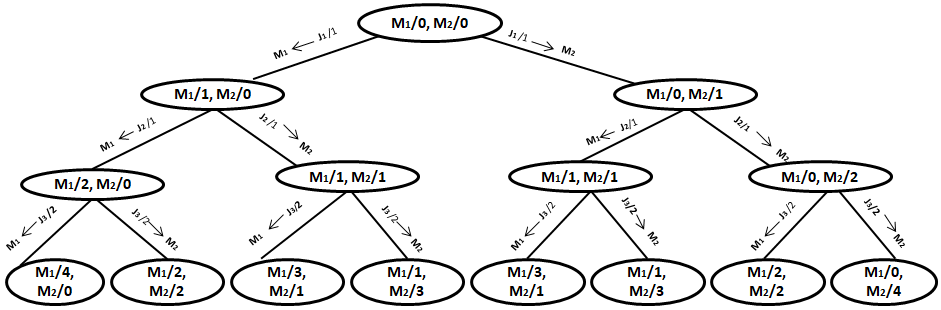}
\caption{Weighted Scheduling Solution Space Tree for 3 Jobs and 2 Processors}
\label{Weightedschedulingsolutionspacetree}
\end{figure}

Before presenting the combinatorial structure of MPSP, we define the following basic terms.\\\\
\textbf{Definition 3.3. Level} of a tree defines the positioning of the parent node and its children in the sense that if the parent is at level $b$ ($\geq 0$), then its child nodes are at level $b+1$ assuming that the root node is at level $0$. \\\\
\textbf{Definition 3.4. Leaf} node of a tree is a node with no children, where as a non-leaf node has at least one child node.\\\\
\textbf{Definition 3.5. Depth or Height ($h$)} of a tree can be defined as the maximum level of the tree, where there exists no non-leaf nodes.\\\\
The combinatorial structure represents all possible solutions of a computational problem \cite{Horowitz08}. In the case of MPSP, if we consider one job and $m$ processors, then we have $m$ possibilities to schedule the job. 
Thus, for scheduling $n$ jobs on $m$ processors ($n > m$), we can have at most $m^n$ possible schedules, which is an exponential solution space.\\
For a basic understanding, we explore the combinatorial structure for the restricted case of the MPSP problem, where $m=2$. We present the assignment of jobs to processors as configurations through the construction of a \textit{Scheduling Solution Space Tree (SSST)} as shown in Figure \ref{Schedulingsolutionspacetree}.
\\
In a SSST, each node $N$ represents a $n$-tuple of the type $J_i/M_j$, which shows the assignment of job $J_i$ to processor $M_j$. Each tuple in the root node represents the \textit{initial configuration} $J_i/\epsilon$, where none of the jobs has been assigned to any of the machines. The tuple $J_i/\epsilon$ changes to the tuple $J_i/M_j$ through an arc with action represented by $J_i\rightarrow M_j$ when job $J_i$ is assigned to processor $M_j$. The leaf nodes are represented by tuples of the type $J_i/M_j$ which shows the assignment of all jobs to some machines.  Let's consider the case, where $m=2$ with $M=\{M_1, M_2\}$ and $n=3$ with $J=\{J_1, J_2, J_3\}$. Let the initial configuration be represented by the root node $N_0$ as a $3$-tuple, where $N_0 : (J_1/\epsilon, J_2/\epsilon, J_3/\epsilon)$. For scheduling job $J_1$ either on processor $M_1$ or on processor $M_2$,  two child nodes (let $N_1$ and $N_2$) are created through arcs $J_1\rightarrow M_1$ and $J_1\rightarrow M_2$ in the SSST to capture two possible new configurations, where $N_1 : (J_1/M_1, J_2/\epsilon,  J_3/\epsilon)$ and $N_2 : (J_1/M_2,  J_2/\epsilon,  J_3/\epsilon)$ respectively. Thus, two child nodes are inserted in the SSST to capture two possible assignments. Recursively, new nodes are created from each internal node by exploring two alternative schedules for each unscheduled job. At level $i$ of the SSST, where $1\leq i\leq n$, all possible assignments of $i^{th}$ job are represented in $2^i$ nodes (of course, we do not consider the root node which resides at level $0$). Each leaf node represents a schedule $S$ with configuration $(J_1/M_j, J_2/M_j, J_3/M_j)$, where $j \in \{1, 2\}$.\\\\
\textbf{Weighted Scheduling Solution Space Tree (WSSST).} It is an extension of the SSST data structure. While the SSST showcases the assignment of jobs to processors, the WSSST data structure represents the intermediate and final loads of the processors in all possible schedules. We present the job assignment and updated loads of the processors as configurations through the construction of a WSSST  by considering an instance of the MPSP as shown in Figure \ref{Weightedschedulingsolutionspacetree}.
We consider $2$ processors and a set $J=\{J_1/1, J_2/1, J_3/3\}$ of $3$ jobs, where each job $J_i$ is characterized by its processing time $p_i$.
\\
In a WSSST, each node $N$ represents a $m$-tuple of the type $M_j/l_j$, which shows the value of the load $l_j$ of the corresponding machine $M_j$, where $l_j=\sum_{J_i\in M_j}^{}{p_i}$, the sum of processing times of the jobs assigned to $M_j$. Each tuple in the root node represents the \textit{initial configuration} $M_j/0$ indicating none of the jobs has been assigned to any of the machines. The tuple $M_j/0$ changes to the tuple $M_j/l_j$ through an arc with action represented by $J_i\rightarrow M_j$ when job $J_i$ is assigned to processor $M_j$. The leaf nodes are represented by tuples of the type $M_j/l_j$ reflecting the final load of machine $M_j$, and the makespan $C_{max}$ of the job schedule can be computed as $C_{max}=\max\{l_j\hspace{0.1cm}| \hspace{0.1cm} 1\leq j\leq m\}$.  
\subsection{Our Analytical Results Based on the Characterization of SSST}
Here, we explore and present some of the important properties of the \textit{SSST} and show a non-trivial characterization of the combinatorial solution space of \textit{MPSP}. We begin with the following definitions. \\\\
\textbf{Definition 3.6.} A \textbf{Path} \cite{Horowitz08} in a tree can be defined as the collection of nodes and edges along the way from the root node to one of the leaf nodes. The first node in the collection is the root node and the other nodes and edges are chosen as follows: at each level $b$, where $1\leq b\leq h$, exactly one node is chosen whose parent has already been chosen at level $b-1$ along with the edge that joins the parent at level $b-1$ to its child at level $b$.\\ \\
\textbf{Definition 3.7.} A \textbf{perfect $m$-ary tree} is a $m$-ary tree, where each node has exactly $m$ child nodes and all leaf nodes are at the same level.\\\\
\textbf{Lemma 3.8.} \textit{$N(h)=\frac{m^{h+1}-1}{m-1}$, where $N(h)$ is the total number of nodes of the perfect $m$-ary tree of height $h$.}\\\\
\textbf{Proof}  Let $N^b$ be the number of nodes at height $b$, where $0\leq b\leq h$. Clearly, $N^0=1=m^0$ as only a single node resides at the root. By \textit{Definition 3.3}, we have $N^1$=$m$, $N^2$=$m^2$ and so on. In general, we can write $N^h$=$m^h$. Therefore, we have $N(h)$=$m^0+m^1+m^2+...+m^{h-1}+m^h$=$\frac{m^{h+1}-1}{m-1}$.\\
We prove the \textit{Lemma 3.8} by method of induction as follows.\\
\textit{Induction Basis:} If $h=0$, then we have $N(h)=\frac{m^{0+1}-1}{m-1}=1=m^0$. Thus, $N(0)$ holds true for height $h=0$.\\
\textit{Induction Hypothesis:} Let us assume $N(k)=\frac{m^{k+1}-1}{m-1}$ is true for height $h=k$.\\
\textit{Inductive Step:} We have to show that $N(k+1)$ is true for $h=k+1$, i.e., $N(k+1)=\frac{m^{k+2}-1}{m-1}$. We can construct a \textit{perfect $m$-ary tree} of height $k+1$ by adding $m^{k+1}$ nodes to $N(k)$ at level $k+1$. Hence, $N(k+1)=N(k)+m^{k+1}$=$\frac{m^{k+2}-1}{m-1}$. Therefore, $N(k+1)$ holds true for $h=k+1$ and $N(h)$ is true for all $h\geq 0$. \hfill\(\Box\)\\\\
\textbf{Theorem 3.9.} \textit{If SSST for MPSP with $m$ processors and $n$ jobs is a perfect $m$-ary tree, then}\\
\hspace*{1.2cm} (a) $h$=$n$ and $|S|$=$m^n$.\\
\hspace*{1.2cm} (b) $|S_p|$=$\frac{m^n-m}{m-1}$.\\
\hspace*{1.2cm} (c) $|S_e|=m^n-m$\\\\
\textbf{Proof} Let us consider the set $J=\{J_1, J_2, J_3, \ldots, J_n\}$ of $n$ jobs  to be scheduled on $m$ processors.\\
\textbf{(a)} To represent all possible options of assignment of each job $J_i$ to one of the $m$ processors, the SSST creates $m$ child nodes for each node at level $i-1$, where $i\geq 1$. Let $N^i$ be the number of nodes at level $i$. Clearly, $N^0=1$, $N^1=1\cdot m=m$, $N^2=m\cdot m=m^2$ and so on. Without loss of generality, we can write $N^i=m\cdot N^{i-1}$, where each node represents a distinct schedule for jobs $J_1, J_2, \ldots, J_i$. Thus, at level $i=n$, \textit{SSST} represents distinct schedules for the set $J$ of jobs by $m^n$ leaf nodes. This implies $|S|=m^n$ and $N(n)=\frac{m^{n+1}-1}{m-1}$. Therefore, by \textit{Lemma 3.8}, we can conclude that the \textit{SSST} for \textit{MPSP} with $m$-processor and $n$ jobs is a \textit{perfect $m$-ary tree} of height $h=n$.\\\\
\textbf{(b)} It is clear from the construction of the SSST that each level $i$ represents $m^i$ \textit{partial schedules} for a set $J^{'}=\{J_1, J_2, \ldots, J_i\}$ of jobs, where $1\leq i\leq n-1$ and $J^{'}\subset J$. Therefore, we have $|S_p|=m^1+m^2+ \ldots +m^{n-1}$=$\frac{m^n-m}{m-1}$.\\\\
\textbf{(c)} Let us consider a \textit{path} in \textit{SSST} from root to one of the leaf nodes, such that at each level $i$, where $1\leq i\leq n-1$, the job $J_i$ is assigned to same machine $M_j$. Clearly, for a \textit{$m$-ary SSST}, there are $m$ leaf nodes representing the assignments of all jobs to any one of the $m$ machines. This implies, the remaining leaf nodes represent the schedules, where at least one job is assigned to all $M_j$, where $1\leq j\leq m$. Therefore, we have $|S_e|=m^n-m$.\\
This completes the proof of Theorem 3.9. \hfill\(\Box\)   
\section{Our Results on Hardness of MPSP}
In this section, we prove that \textit{MPSP} is $\mathcal{NP}$-complete by simulating a Non-deterministic Turing Machine ($\mathcal{NTM}$) through our proposed SSST and reducing an instance of MPSP from an instance of well-known Partition problem. Our objective is to present exhaustive and refined proof sketches of $\mathcal{NP}$ and $\mathcal{NP}$-hardness for a better understanding of the hardness of the MPSP problem. 
\subsection{MPSP $\in \mathcal{NP}$}
Here, we first prove $MPSP\in \mathcal{NP}$ by mapping the construction of SSST as a $\mathcal{NTM}$. Alternatively, we develop a polynomial-time verifier by combinely using SSST and WSSST data structures.\\\\
\textbf{Theorem 4.1.} \textit{$MPSP\in \mathcal{NP}$.}\\\\
\textbf{Proof} We consider an instance of \textit{MPSP}, where $m=2$. Let us assume there exists an encoding that can represent the MPSP with at least $n$ symbols. Without loss of generality, let the set $J=\{J_1, J_2, J_3, \ldots, J_n\}$ denotes the input symbols and the set $M=\{M_1, M_2\}$ represents the output alphabet of the $\mathcal{NTM}$. Let the set $N=\{\{N_0\}\cup\{N_1, N_2, \ldots, N_{2^{n+1}-2}\}\}$ represents the nodes of the SSST, where $N_0$ is the root node. We now can couch the SSST as a state transition diagram generated by the $\mathcal{NTM}$, where the root node represents the initial state of the 1-tape $\mathcal{NTM}$ and each node represents a state with configuration resulting due to the transition function, which is defined as $\delta: N X J\rightarrow N X M_j X R$, where $j\in \{1, 2\}$ and $R$ represents the right-hand side move off the tape header.\\
It is clear by \textit{Theorem 3.9} that any scheduling certificate representing the assignment of jobs can be verified by traversing the SSST from root to one of the leaf nodes in polynomial time bounded by the height of the tree.\\
After representing all schedules at the leaf nodes of SSST, we now consider the WSSST, where each node ($N_x$) is assigned with a \textit{weight} to represent the current loads of the machines. Therefore, each leaf node now represents a scheduling certificate with a makespan value.  It is assumed that the value of the optimum makepsan $C_{OPT}=\frac{\sum_{i=1}^{n}{p_i}}{m}$. We consider $C_{OPT}$  as a reference to compare with the weight of any leaf node. Let $W$ be a weight function from $N$ to non-negative integers, which represents the value of the makespan at any $N_x$. We consider $W(N_x)=\max_{1\leq j\leq 2}^{}{l_j}$, where $l_j=\sum_{J_i\in M_j}^{}{p_i}$ and $1\leq x\leq 2^{n+1}-2$. W.l.o.g., we assume $W(N_0)=0$. The optimality of any scheduling certificate can be verified in polynomial time by comparing the weight of the leaf node with the value of $C_{OPT}$. \\
Therefore, it is proved that $MPSP\in \mathcal{NP}$.\hfill\(\Box\)   
\subsection{SSST as a Polynomial-time Verifier}
The $\mathcal{NP}$ class covers all decision problems for which the given solutions are verifiable in polynomial time \cite{Garey90}. According to the interactive proof procedures of Goldwasser et al. \cite{Goldwasser89}, the verification process considers a \textit{prover} and a \textit{verifier}. The \textit{prover} provides a yes/no answer to the decision version of a computational problem with an explanation in the form of a solution certificate. The verifier validates the answer in polynomial time and concludes whether the provided answer is correct or not. For a proper validation, the verifier must know all feasible solutions and the optimal one. Let us consider an instance of the MPSP problem with $n$ jobs, corresponding processing times, and the value of the optimum makespan $C_{OPT}$. We use the SSST and WSSST together as a verifier for the MPSP problem.\\\\
\textit{Decision version of MPSP.} Is there a schedule $S$ for a given list of $n$ jobs on $m$ identical processors such that $C_{max}\leq C_{OPT}$? \\
Suppose the True answer is "yes".\\
\textit{Prover.} It provides a schedule $S$ with the value of $C_{max}$.\\
\textit{Verifier.} Given the schedule $S$ and the value $C_{max}$, the verifier must validate the following in the polynomial time.
\begin{enumerate}
\item \textit{The given $S$ is a valid schedule of a list of $n$ jobs on $m$ processors}.\\ This can be verified by traversing the SSST from root to a leaf node, and the time required to complete the traversal is bounded in polynomial time by the height of the SSST.
\item \textit{The $C_{max}\leq C_{OPT}$.}\\  For a given instance of the $MPSP$, we can know the actual value of the optimum makespan by comparing the values of the leaf nodes of the WSSST. After knowing the exact value of $C_{OPT}$, the verification of $C_{max}\leq C_{OPT}$ requires a constant time.    \\
\end{enumerate}
Although it has been considered in the theory the value $C_{OPT}=\frac{1}{m}\cdot \sum_{i=1}^{n}{p_i}$, it does not reflect the actual value of $C_{OPT}$ in all instances of the problem. The theoretical value of $C_{OPT}$ is impractical for real-life instances. Therefore, WSSST would be handy in deriving the actual value of $C_{OPT}$ in practical scenarios.\\
The SSST is an implicit representation, meaning that the nodes and the branches are considered implicit objects in the computer's memory. One can algorithmically generate the components of the SSST  for a given instance of the problem. A typical SSST is a larger structure to represent in the memory. The nodes of the SSST can be generated while exploring a path from the root to a leaf node to verify a scheduling solution certificate and are discarded subsequently. Therefore the SSST data structure could be a global polynomial-time verifier for any instance with a solution certificate of the MPSP problem.
\subsection{MPSP $\in \mathcal{NP}$-hard}
Here, we first show the $\mathcal{NP}$-hardness of the MPSP problem by a polynomial-time reduction from the well-known Partition problem. Then, we develop a new reduction framework to design the first non-deterministic polynomial time algorithm for the MPSP problem.\\\\
\textbf{Theorem 4.3.} \textit{MPSP $\in \mathcal{NP}$-hard.}\\\\
\textbf{Proof} We start with defining the \textit{Partition Problem (PP)} and show the polynomial time reduction of an instance of the \textit{MPSP} from an instance of the \textit{PP}.\\
\textit{Partition Problem:} Given a set $A=\{a_1, a_2, \ldots, a_{n-1}, a_n\}$ of $n$ elements, where each $a_i$ has a non-negative weight of $w_i$ and the total weight of set $A$ is $W(A)=\sum_{a_i\in A}^{}{w_i}$.\\
\textit{Decision Version of PP:} Can we have partitions of set $A$ into $r$ disjoint sets such that $W(A_1)=W(A_2)= \ldots =W(A_r)=\frac{W(A)}{r}$?\\
\textit{Instance of PP:} We have an instance of \textit{PP} by considering $r=2$. \\
We now define the \textit{decision version} of $2$-PP as follows:\\
Can we have a subset $A^{'}\subset A$ such that $\sum_{a_i\in A^{'}}^{}{w_i}=\sum_{a_i\in A-A^{'}}^{}{w_i}=\frac{W(A)}{2}$?\\
\textit{Instance of MPSP:} We consider an instance of \textit{MPSP} with $m=2$.\\
\textit{Polynomial Transformation of \textit{$2$-PP} to $2$-Processor Scheduling Problem (\textit{$2$-PSP}):} Let us consider $D=W(A)$ and $C_{OPT}=\frac{\sum_{i=1}^{n}{p_i}}{2}$. We now show the equivalence between the input parameters of \textit{$2$-PP} and \textit{$2$-PSP} as follows: $J_i=a_i$ and $p_i=w_i$ for $1\leq i\leq n$; $m=2$, jobs assigned to machine $M_1$ are equivalent to the elements belong to set $A^{'}$ and the jobs assigned to machine $M_2$ are equivalent to the elements belong to set $A-A^{'}$; we consider $C_{max}=\max\{W(A^{'}), W(A-A^{'})\}$.\\
Clearly, the transformation is bounded by a polynomial in the length of the inputs represented as the set elements or jobs.\\
We now can claim that \textit{$2$-PSP} has a schedule $S$ such that $C_{max}\leq C_{OPT}=\frac{D}{2}$ if and only if there exists a subset $A^{'}\subset A$ such that $W(A^{'})=W(A-A^{'})=\frac{D}{2}$ for the positive number $D$. \hfill\(\Box\)
\subsection{A Framework for designing non-deterministic Polynomial-time Algorithm} 
\begin{figure}[!htbp]
\centering
\includegraphics[scale=.83]{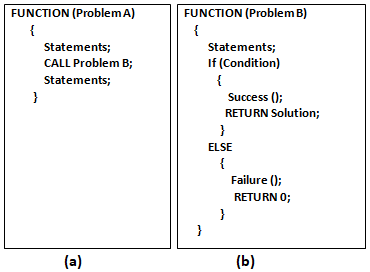}
\caption{Sample Subroutines for \hspace*{0.1cm} (a)\hspace*{0.1cm} Problem A \hspace*{0.1cm}(b)\hspace*{0.1cm} Problem B}
\label{ageneralreductionframework}
\end{figure}

Recall that the complexity class $\mathcal{NP}$ defines a set of problems for which we do not have deterministic polynomial-time algorithms, but the class accepts the existence of non-deterministic polynomial-time algorithms for these problems \cite{Garey90}. We develop a general framework for designing a non-deterministic polynomial-time algorithm for the MPSP problem by using the principle of \textit{reduction} discussed in Section 2.1. Let us consider that we have a computational problem $B$ and a well-known $\mathcal{NP}$-hard problem $A$. If an instance of problem $A$ is reduced $(\propto)$ to an instance of problem $B$, we can solve problem $A$ in polynomial time if and only if there exists a polynomial-time solution for problem $B$ and vice-versa. Moreover, we can develop a polynomial-time algorithm ${Alg}_A$ for $A$ by using the polynomial-time algorithm designed for problem $B$ as a subroutine for ${Alg}_A$ \cite{Garey90}. By considering this reduction principle, we present sample subroutines for problem $A$ and $B$ in Figure \ref{ageneralreductionframework}(a) and Figure \ref{ageneralreductionframework}(b). The function Success()  in the subroutine for problem $B$ indicates the existence of the optimal output, and the Failure() function represents the non-existence of the desired outcome. If the subroutine for $B$ runs in polynomial time, then the subroutine for $A$ will also run in polynomial time. Next, we develop a non-deterministic polynomial-time algorithm for the MPSP problem using the general reduction framework.  
\subsection{A Non-deterministic Polynomial-time Algorithm : Magic Scheduling}
A non-deterministic algorithm makes random choices for each input in different runs resulting in different outputs \cite{Cormen09}. In a non-deterministic algorithm, we may have some statements or functions whose operations are not defined clearly in the context of the problem. However, these statements can guide in exploring and verifying the optimal solution for the problem. For example, if a solution certificate to an instance of the $\mathcal{NP}$-complete problem is given, the algorithm can verify whether the given solution is an optimal one or not. Hence, a non-deterministic algorithm works as a verifier than a solver. We design a non-deterministic algorithm named \textit{Magic Scheduling (MS)} by using our proposed reduction framework. We present the pseudocode of the algorithm MS as follows.
\begin{algorithm}[!ht]
\caption{MS}
\begin{algorithmic}[1]
\STATE Initially $l_1=l_2=0$\\
\STATE PARTITION($X, w_i$)\hspace*{0.8cm} // a set $X$ with $n$ elements, each one having the corresponding weight $w_i$\\
\STATE \hspace*{0.2cm} BEGIN\\
\STATE \hspace*{0.5cm} $P(A, A^{'}) \leftarrow$ SHEDULE $(2, n, p_i, C_{OPT})$\hspace*{0.8cm}// $C_{OPT}=\frac{1}{2}\cdot \sum_{i=1}^{n}{p_i}$ \\
\STATE \hspace*{0.2cm} END\\
\STATE SCHEDULE($m, n, p_i, C_{OPT}$)\\
\STATE \hspace*{0.2cm} BEGIN\\
\STATE \hspace*{0.5cm} $P(a, a^{'})\leftarrow$ SELECT$\_$PARTITION() \hspace*{0.5cm}// non-deterministic function\\
\STATE \hspace*{0.5cm} $a=\{J_i\hspace*{0.1cm}|\hspace*{0.1cm} J_i\in M_1\}$ \hspace*{3.1cm}// assigns jobs to machine $M_1$\\
\STATE \hspace*{0.5cm} $a^{'}=\{J_i\hspace*{0.1cm}|\hspace*{0.1cm} J_i\in M_2\}$ \hspace*{3.0cm}// assigns jobs to machine $M_2$\\
\STATE \hspace*{0.5cm} $l_1=\sum_{J_i\in M_1}^{}{p_i}$, \hspace*{0.1cm} and \hspace*{0.1cm} $l_2=\sum_{J_i\in M_2}^{}{p_i}$\\
\STATE \hspace*{0.5cm} $C_{max}\leftarrow \max\{l_1, l_2\}$\\
\STATE \hspace*{0.5cm} IF ($C_{max}= C_{OPT}$)\\
\STATE \hspace*{0.7cm} Success () \hspace*{3.8cm}// there exists an optimum schedule \\
\STATE \hspace*{0.7cm} Return $P(a, a^{'})$ \\
\STATE \hspace*{0.5cm} ELSE\\
\STATE \hspace*{0.7cm} Failure () \hspace*{3.8cm}// there does not exist an optimum schedule\\
\STATE \hspace*{0.7cm} Return 0\\
\STATE \hspace*{0.2cm} END
\end{algorithmic}
\end{algorithm} \\
Algorithm MS runs in polynomial time at the current state, and its worst-case time complexity depends on the function SELECT$\_$PARTITION (). One can observe that in algorithm MS, the function SELECT$\_$PARTITION is a non-deterministic function, which partitions a list of $n$ jobs into two subsets $A$ and $A^{'}$. The jobs belonging to subset $A$ are assigned to machine $M_1$, and the jobs belong to set $A^{'}$ to machine $M_2$. On the contrary, we do not know how the SELECT$\_$PARTITION function selects jobs for each subset. However, if we can determine the exact working principle, then we can obtain an optimal solution for the MPSP problem, and using the same, we can also solve the Partition problem.  \\
Nevertheless, if a solution to an instance of MPSP is given, then the SELECT$\_$PARTITION function can use the SSST to select an appropriate partition from the leaf nodes to verify whether the solution certificate is the optimal one or not. Therefore, the SSST data structure with our proposed non-deterministic polynomial-time algorithm MS constitutes a computational framework to show MPSP $\mathcal{NP}$-complete. 
\section{Hardness of Multiuser Multiprocessor Scheduling Problem}
In the MPSP problem, we do not explicitly consider the sources of job requests and assume that all jobs arrive from a single source only. MUMPSP is a practically significant variant of MPSP, where \textit{user} is considered as an additional input parameter to represent the source of job requests. Here, each \textit{user} submits a set of jobs to the scheduler and aims for obtaining the optimal makespan. In the MUMPSP problem with $k (\geq 1)$ users, the set $J$ of jobs considered in MPSP can be partitioned into $k$ disjoint subsets, where each subset represents the jobs of a single user, and the scheduler must meet the objective of each user. This scheduling problem often arises in high-performance computing systems such as supercomputers, cloud servers, robotic controllers, clusters, and grids \cite{Saule09}.
\subsection{Multiuser Multiprocessor Scheduling Problem}
We formally define the MUMPSP by presenting the inputs, output, objective and assumptions of the problem as follows. 
\begin{itemize}
\item Inputs: \begin{itemize}
\item Given, $M=\{M_1, M_2, \ldots, M_m\}$ is the set of $m$ identical processors.
\item Let $U_r$ is an \textit{user}, where $1\leq r\leq k$ and $L_{r}$ denote a distinct set of jobs of respective $U_r$, where $L_{r}=\{J^{r}_i|1\leq i\leq n_r\}$ such that $\sum_{r=1}^{k}{n_r}=n$
\item Let $p^{r}_i$ is the processing time of job $J^{r}_i$, where $p^{r}_i\geq 1$.
\end{itemize}
\item Output: Generation of a Schedule($S$), representing for each \textit{user} $U_r$, the \textit{makespan} $C^{r}_{max}=\max\{C^{r}_i|1\leq i\leq n_r\}$, where $C^{r}_i$ is the completion time of job $J^{r}_i$ such that $(t(S)+p^{r}_i)\leq C^{r}_i\leq C^{r}_{max}$, where $t(S)$ is the starting time of the schedule $S$.
\item Objective: Minimization of $C^{r}_{max}$, \hspace*{0.1cm}  $\forall U_r$.
\item Assumptions: \begin{itemize}
\item Jobs are submitted from $k$ users.
\item The jobs of each user are independent of each other as well as with the jobs of the other users. 
\item No job can be splitted and preempted in the middle of its execution.
\end{itemize}
\end{itemize}
\subsection{Our Results on Hardness of MUMPSP}
We prove the MUMPSP problem $\mathcal{NP}$-complete by a polynomial-time reduction from the MPSP problem. \\\\
\textbf{Theorem 5.2.} \textit{MUMPSP $\in \mathcal{NP}$-complete.} \\\\
\textbf{Proof} We first show $2$-PSP $\propto$ MUMPSP.\\
\textit{Problem:} $2$-PSP:\\
\textit{Instance:} Given $m=2$, and a list $J=\{J_1/p_1, J_2/p_2, J_3/p_3, \ldots, J_n/p_n\}$ of $n$ jobs, where $p_i\geq 1$.\\
\textit{Decision version:} Is there a schedule $S$ of the list $J$ of $n$ jobs on $2$ processors such that $l_1=l_2=\frac{1}{2}\cdot \sum_{i=1}^{n}{p_i}=D$? \\
We now have to construct an instance $I^{'}\in$ MUMPSP from an instance $I\in$  $2$-PSP.\\
Suppose, the number of users $k=1$ for MUMPSP, and we denote the instance as $1$-UMPSP. Then, $I^{'}$ can be constructed by letting $n_1=n$, $p^{1}_{i}=p_i$ for $1\leq i\leq n$, and $C^{1}_{OPT}=\frac{1}{2}\cdot \sum_{i=1}^{n}{p^{1}_{i}}=D$.\\
Clearly, the construction of $I^{'}$ requires total $n+2$ assignments, and $n$ additions to compute $C^{1}_{OPT}$. Therefore, the mapping of the instance $I$ into $I^{'}$ requires the time at most $O(n)$.\\
To ensure completeness, we have to show that the answer to the decision version of the instance $I\in 2$-PSP is 'yes' if and only if for the instance $I^{'}\in 1$-UMPSP the answer is 'yes'. Let $S$ be the optimal schedule for $I$ such that $l_1=l_2=D$. Let $A_1=\{J_i\hspace*{0.1cm}|\hspace*{0.1cm} Jobs \hspace*{0.1cm} assigned \hspace*{0.1cm} to \hspace*{0.1cm} M_1\}$ and $A_2=\{J_i\hspace*{0.1cm}|\hspace*{0.1cm} Jobs \hspace*{0.1cm} assigned \hspace*{0.1cm} to \hspace*{0.1cm} M_2\}$. We now can construct a solution for $I^{'}$ by scheduling the corresponding job $J^{1}_{i}=J_i\in A_1$ on $M_1$ and the job $J^{1}_{i}=J_i\in A_2$ on $M_2$. Let $S^{'}$ be the resulting schedule for $I^{'}$. By definition of optimal schedule $S$, we have $\sum_{J^{1}_{i}\in M_1}^{}{p^{1}_{i}}=\sum_{J^{1}_{i}\in M_2}^{}{p^{1}_{i}}=\frac{1}{2}\cdot \sum_{i=1}^{n}{p^{1}_{i}}=D$.\\
Clearly, given $S$, the schedule $S^{'}$ is constructed in $O{n}$ time. Similarly, given the optimal schedule $S^{'}$ for $1$-UMPSP, we can obtain in polynomial time an optimal schedule $S$ by mapping the same assignment procedure. This completes the reduction part to show that  $MUMPSP$ is $\mathcal{NP}$-hard. Now we establish an analogy to show the $\mathcal{NP}$-completeness of the problem. \\
As we know that the \textit{MPSP} problem is a special case of the \textit{MUMPSP} with number of users $k=1$. Thus, by \textit{Property 2} of the complexity classes discussed in Section 2.1, we conclude that \textit{MUMPSP} is $\mathcal{NP}$-complete. \hfill\(\Box\)
\section{Open Problems}
We present some interesting open problems for future investigation.
\begin{itemize}
\item Can the \textit{SSST} data structure be used to explore the combinatorial structure of a class of $\mathcal{NP}$-complete problems? If yes, then define mappings through the SSST to show the equivalence among the $\mathcal{NP}$-complete problems.
\item Can we have an alternate and simple $\mathcal{NP}$-hard reduction of the MUMPSP problem from any of the well-known $\mathcal{NP}$-complete problems other than the MPSP problem? 
\item What is the mathematical formulation to achieve the practically significant optimum value of the makespan $C_{OPT}$ for all instances of the MPSP in general?
\item Can SSST be extended as a \textit{disjoint-set data structure} to explore the combinatorial structure of MUMPSP? If yes, then find the optimum bound for $C^{r}_{max}$, $\forall r$.
\item  Is MUMPSP with $k=m$ polynomially bounded?
\end{itemize}



\begin{thebibliography}{2}
%
\bibitem{Aaronson16}
Aaronson S. $P=? \NP$, \emph{Open Problems in Mathematics}, 2016, 1--122.
%
\bibitem{Aho74}
Aho A. V., Hopcroft J. E., Ullman J. D., The Design and Analysis of Computer Algorithms,  \emph{Addison-Wesley}, Reading, MA, 1974.
\bibitem{Ambuhl11}
Ambuhl C., Mastrolilli M., Mutsanas M., Svensson O., On the approximability of single machine scheduling with precedence constraints, \emph{Mathematics of Oper. Res.}, \textbf{36}, 2011, 653--669.
\bibitem{Bellenguez15}
Bellenguez-Morineau O., Chrobak M., D\"urr C., Prot D., A note on NP-Hardness of preemptive mean flow-time scheduling for parallel machines, \emph{Journal of Scheduling}, \textbf{18}, 3, 2015, 299--304.
\bibitem{Bevern16}
Bevern R. V., Bredereck R., Bulteau L., Komusiewicz C., Tulmon N., Woeginger G. J., Precedence-constrained scheduling problems parameterized by partial order width, \emph{DOOR}, 2016, 105--120.
\bibitem{Blazewicz01}
Blazewicz J., Ecker K., Kis T., Tanas M., A note on the complexity of scheduling coupled tasks on a single processor, \emph{J. Braz. Comput. Soc.}, \textbf{7}, 2001, 23--26.
\bibitem{Bodlaender95}
Bodlaender H. L., Fellows M. R., $W[2]$-hardness of precedence constrained $K$-processor scheduling, \emph{Oper. Res. Lett.}, \textbf{18}, 1995, 93--97.
\bibitem{Bruno74}
Bruno J., Coffman E. G.,Sethi R., Scheduling independent tasks to reduce mean finishing time, \emph{Comm. ACM}, {\bf 17}, 1974, 382--387.
\bibitem{Cook70}
Cook S. A., The complexity of theorem proving procedures,  \emph{Proceedings of the 3rd ACM Conference on Theory of Computing}, May, 1970, 151--158.
\bibitem{Cormen09}
Cormen T. H., Leiserson C. E., Rivest R. L., Stein C., Introduction to Algorithms, $3^{rd}$ Edition, MIT Press, 2009, ISBN: 978-0-262-03384-8.
\bibitem{Davies21}
Davies S., Kulkarni J., Rothvoss T., Sandeep S.,  Tarnawski J., Zhang Y., On the hardness of scheduling with non-uniform communication delays, \emph{CoRR abs/2109.01064}, 2021.
\bibitem{Garey74}
Garey M. R., Johnson D. S., Complexity results for multiprocessors scheduling under resource constraints, \emph{Proceedings of the 8th Annual Princeton Conf. on Inf. Sci. and Systems}, August, 1974, 384--393.
\bibitem{Garey90}
Garey M. R., Johnson D. S., Computers and intractability : A guide to the theory of  NP-completeness, $2^{nd}$ Edition, W. H. Freeman $\&$ Co., 1990, New York, USA.
\bibitem{Garg07}
Garg N., Kumar, A., Pandit V., Order scheduling models: Hardness and algorithms, \emph{FSTTCS}, 2007, 96--107.
\bibitem{Goldwasser89}
Goldwasser S., Micali S., Rackoff C., \emph{The knowledge complexity of interactive proof systems, SIAM Journal of Comput.}, \textbf{18}, 1, 1989, 186--208.
\bibitem{Graham95}
Graham R. L., Groetschel M., Lovasz L., eds., Handbook of Combinatorics, Vol. 1, \emph{Elsevier (North-Holland)}, ISBN 0-262-07169-X, 1995.
\bibitem{Horowitz08}
Horowitz E., Sahni S., Anderson-Freed S., Fundamentals of Data Structures in C, $2^{nd}$ Edition \emph{Universities Press}, 2008.
\bibitem{Karp72}
Karp R. M., Reducibility among combinatorial problems, \textbf{TR3}, Department of Computer Science, University of Califormia at Berkeley, April, 1972.
\bibitem{Lenstra77}
Lenstra J. K., Rinnooy Kan A. H. G., Brucker P., Complexity of machine scheduling problems, \emph{Annals of Discrete Mathematics}, {\bf 1}, 1977, 343--362.
\bibitem{McNaughton59}
McNaughton R., Scheduling with deadlines and loss functions,  \emph{Management Science}, {\bf 6}, 1959, 1--12.
\bibitem{Mnich15}
Mnich M., Wiese A., Scheduling and fixed-parameter tractability, \emph{Math. Program., Ser. B}, \textbf{154}, 2015, 533--562.
\bibitem{Saule09}
Saule E., Trystram D., Multi-users scheduling in parallel systems, \emph{Proceedings of the IEEE International Parallel and Distributed Processing Symposium},  Washington DC, USA, May, 2009, 1--9.
\bibitem{Svensson11}
Svensson O., Hardness of precedence constrained scheduling on identical machines, \emph{SIAM J. Comput.}, \textbf{40}, 2011, 1258--1274.
\bibitem{Ullman75}
Ullman J. D., NP-Complete Scheduling Problems, \emph{Journal of Computer and System Sciences}, {\bf 10}, 1975, 384--393.
\bibitem{Zhang18}
Zhang An., Chen Y., Chen L., Chen G., On the NP-hardness of scheduling with time restrictions, \emph{Discret. Optim.}, \textbf{28}, 2018, 54--62.
\end{thebibliography}

\end{document}